\documentclass[AMA,STIX1COL]{WileyNJD-v2}

\newcommand\BibTeX{{\rmfamily B\kern-.05em \textsc{i\kern-.025em b}\kern-.08em
T\kern-.1667em\lower.7ex\hbox{E}\kern-.125emX}}

\articletype{Article Type}%

\received{<day> <Month>, <year>}
\revised{<day> <Month>, <year>}
\accepted{<day> <Month>, <year>}

\begin{document}
\title{High-temperature and Abrasion Resistant Selective Solar Absorber under Ambient Environment}

\author[1]{Yanpei Tian}
\author[2]{Lijuan Qian}
\author[1]{Xiaojie Liu}
\author[3]{Alok Ghanekar}
\author[4,5]{Jun Liu}
\author[4]{Thomas Thundat}
\author[2]{Gang Xiao}
\author[1]{Yi Zheng*}

\authormark{Tian \textsc{et al}}

\address[1]{\orgdiv{Department of Mechanical and Industrial Engineering}, \orgname{Northeastern University}, \orgaddress{\state{Boston, MA}, \country{USA}}}

\address[2]{\orgdiv{Department of Physics}, \orgname{Brown University}, \orgaddress{\state{Providence, RI}, \country{USA}}}

\address[3]{\orgdiv{Artech LLC}, \orgname{Morristown}, \orgaddress{\state{NJ}, \country{USA}}}

\address[4]{\orgdiv{Department of Chemical and Biological Engineering}, \orgname{University at Buffalo, The State University of New York}, \orgaddress{\state{Buffalo, NY}, \country{USA}}}

\address[5]{\orgdiv{RENEW Institute, University at Buffalo}, \orgname{University at Buffalo, The State University of New York}, \orgaddress{\state{Buffalo, NY}, \country{USA}}}

\corres{*Yi Zheng,  \email{y.zheng@northeastern.edu}}

\presentaddress{Present address}

\abstract[Abstract]{Selective solar absorbers (SSAs) with high performance are the key to concentrated solar power systems. Optical metamaterials are emerging as a promising strategy to enhance selective photon absorption, however, the high-temperature resistance (> 500$^{\circ}$C) remains as one of the main challenges for their practical applications. Here, a multilayered metamaterial system (Al$_{2}$O$_{3}$/W/SiO$_{2}$/W)  based on metal-insulator-metal (MIM) resonance effect has been demonstrated with high solar absorptance over 92\%, low thermal emittance loss below 6\%, and significant high-temperature resistance: it has been proved that the optical performance remains 93.6\% after 1-hour thermal annealing under ambient environment up to 500$^{\circ}$C, and 94.1\% after 96-hour thermal cycle test at 400$^{\circ}$C, which is also confirmed by the microscopic morphology characterization. The spectral selectivity of fabricated SSAs is angular independent and polarization insensitive. Outdoor tests demonstrate that a peak temperature rise (193.5$^{\circ}$C) can be achieved with unconcentrated solar irradiance and surface abrasion resistance test yields that SSAs have a robust resistance to abrasion attack for engineering applications.
}


\jnlcitation{\cname{%
\author{Williams K.}, 
\author{B. Hoskins}, 
\author{R. Lee}, 
\author{G. Masato}, and 
\author{T. Woollings}} (\cyear{2016}), 
\ctitle{A regime analysis of Atlantic winter jet variability applied to evaluate HadGEM3-GC2}, \cjournal{Q.J.R. Meteorol. Soc.}, \cvol{2017;00:1--6}.}
\maketitle
\section{Introduction}\label{sec1}
\footnotetext{\textbf{Abbreviations:} SSAs, Selective solar absorbers}
Solar energy is abundant, striking our earth at a rate of 90,000 TW, which is 5,000 times of our current global power consumption \cite{abbott2010keeping}. The conversion of solar energy is considered as a promising approach to address the energy and environmental crisis. To date, various solar energy conversion technologies have been developed including photovoltaics\cite{polman2012photonic,lin2012small,schmidt2001self}, photo-thermal methods\cite{tian2013review,garg2012solar,tritt2008thermoelectrics}, and artificial photosynthesis\cite{meyer1989chemical,imahori2003nanostructured,gust2009solar}. Among those technologies, solar thermal power systems, such as concentrated solar power (CSP) via Rankine cycle \cite{mills2004advances}, solar thermoelectric generators (STEGs) \cite{kraemer2011high}, and solar thermophotovoltaics (STPVs) \cite{wang2011novel,tian2018tunable,tian2019near} have attracted increasing attention recently for various environmentally sustainable applications in industrial heating\cite{kaempener2015solar}, air conditioning\cite{lim2017heat}, and electricity generation\cite{kraemer2011high,kraemer2016concentrating}. As a crucial element, selective solar absorbers (SSAs), which are capable of converting solar radiation into heat, have a considerable influence on the overall performance of solar thermal systems. However, it is challenging to achieve high thermal resistance of the SSAs, which operates under a sever high-temperature working environment. Therefore, it is crucial to boost the performance of solar thermal systems by enhancing the sunlight absorption capability by depositing the infrared anti-reflection layer \cite{barshilia2012nanometric,li2018broadband,cao2017high} and improved the resistance under high-temperature environment \cite{wang2018spectrally,wang2018design}. Furthermore, SSAs are required to be omnidirectional and polarization-insensitive in the solar irradiance regime due to the randomly distributed solar irradiation \cite{wang2015highly}. Additionally, the cut-off wavelength of a selective absorber, at which its absorptance spectrum changes steeply, is highly dependent on its operational temperature because of the shifting nature of blackbody radiation according to Wien's displacement law \cite{siegel2001thermal}. However, existing technologies are either relied on high-cost nanofabrications methods or suffer from poor thermal resistance in the ambient environment \cite{cao2014review} and the omnidirectional and polarization insensitivity are scarcely achieved. Hence, SSAs with high-temperature ambient resistance and omnidirectional and polarization insensitivity are highly demanded in solar thermal applications.

Recently, advances in the small-scale fabrications have enabled manipulation of sunlight trapping in micro/nanostructures. Excitation of plasmonic resonance can produce selective and tunable absorption peaks at different wavelengths \cite{zhao2017design}. Simultaneously, the alternation from high UV/visible absorptance to low mid-infrared emittance is sharp as compared with natural materials based SSAs. Abundant metamaterial based selective  absorbers have been investigated, such as 1-D or 2-D surface gratings \cite{ghanekar2018optimal,wang2013perfect,lee2014wavelength,han2016broadband,wang2017broadband,marques2002left,aydin2007capacitor,shchegolkov2010perfect,liu2007plasmon,tao2008metamaterial}, nanoparticles embedded dielectrics \cite{tian2019near,tian2018tunable,ghanekar2017mie,ghanekar2016novel}, crossbar or nano-disk arrays \cite{chen2012dual,Nielsen2012Efficient}, and photonic crystals \cite{stelmakh2013high,celanovic2008two,rinnerbauer2013large,li2015large,jiang2016refractory}. Nevertheless, these absorbers depend on time-consuming nanofabrication technologies, such as photolithography and electron beam lithography, which makes them unrealistic to be scalable-manufactured. Moreover, the high temperature will yield permanent damage to SSAs' spectral selectivity. 

Metal-insulator-metal (MIM) \cite{chirumamilla2016multilayer,langlais2014high,nuru2014heavy} based novel absorbers consist of two metallic layers with a thin dielectric layer sandwiched in between. The incident light is absorbed due to strong optical interaction at the resonant frequencies and converted to heat in lossy metallic nanostructures \cite{zhang2017ultra}. They are feasible to be scale-manufactured with vacuum deposition methods like sputtering\cite{teixeira2001spectrally,fan1977selective,zhang2004high,zhang2000recent}, evaporation \cite{zhang1992new,niklasson1983solar,craighead1981graded} or chemical vapor deposition \cite{gogova1997optical,seraphin1976chemical}. Though high wavelength selectivity has been demonstrated on the multilayer-based SSAs structures, their high-temperature resistance is far from satisfying, mainly due to the lack of materials optimization. The refractory metals or metal oxides, including tungsten (W) \cite{zhang2004high,zhang1997direct}, nickel (Ni) \cite{sathiaraj1990ni,craighead1977optical}, chromium (Cr) \cite{teixeira2001spectrally,fan1977selective}, silica (SiO$_2$) \cite{wang2011optical,esposito2009fabrication,farooq1998high}, alumina (Al$_2$O$_3$) \cite{barshilia2011structure,xinkang2008microstructure,cheng2013improvement,antonaia2010stability}, and chromia (Cr$_2$O$_3$) \cite{nunes2003graded,yin2009direct} have been applied to manufacture SSAs. Specifically, W, Ni, and Cr thin film display high reflectivity in the mid-infrared wavelength region \cite{ghanekar2018optimal}, while maintaining the high absorptance in UV, visible, and the near-infrared regime when they are in the form of nanoparticles \cite{ghanekar2017mie,ghanekar2016novel}. On the other hand, oxides such as SiO$_2$, Al$_2$O$_3$, and Cr$_2$O$_3$ have high-temperature resistance even in the ambient environment. Here, we report a MIM based (W-SiO$_2$-W) SSAs that show high spectral selectivity (solar absorptance/thermal emittance = 13.9) and high thermal efficiency (80.03\% without optical concentration, compared to the state of the art 78\% \cite{wang2015highly}.). The spectral selectivity of fabricated SSAs is angular independent (\textasciitilde75$^{\circ}$, compared to the state of the art \textasciitilde60$^{\circ}$\cite{wang2018design}) and polarization insensitive (0$^{\circ}$ or 90$^{\circ}$.). The optical performance remains 93.6\% and 94.1\% after 1-hour thermal annealing up to 500$^{\circ}$C and 96-hour thermal treatment cycle at 400$^{\circ}$C, respectively. A high stagnation temperature of 193.5$^{\circ}$C was achieved under unconcentrated solar irradiance, which indicates its wide feasibility of low-temperature solar thermal applications without solar concentrators. The sandpaper abrasion robustness tests are also conducted in this work to demonstrate the mechanical abrasion resistance of SSAs' surface.

\section{Results and Discussions} \label{sec2}
\subsection{Energy conversion efficiency analysis of the SSAs}
\begin{figure*}[ht]
\centering
\includegraphics[width=1.0\textwidth]{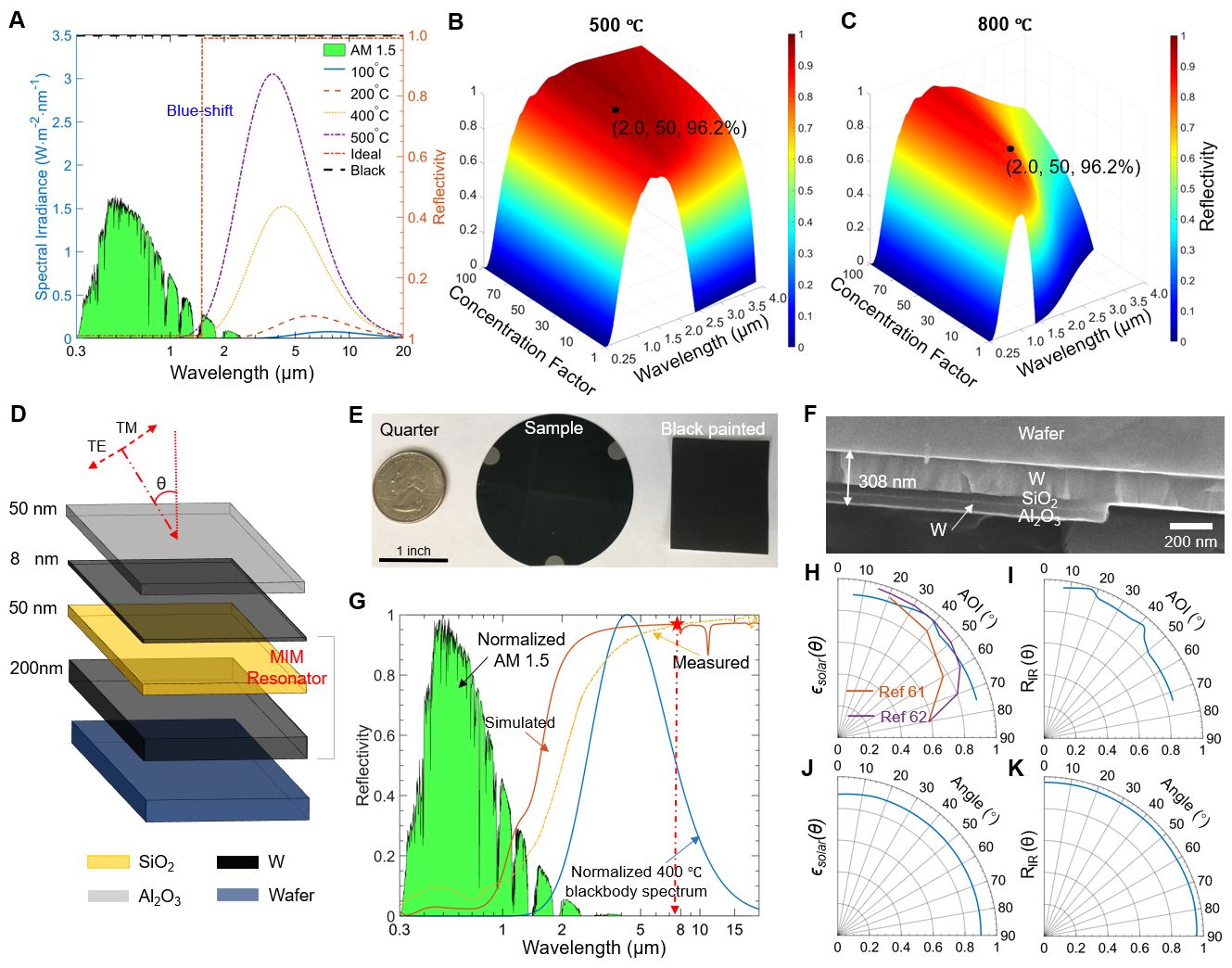}
\caption{\label{fig:sem_variangle} (\textbf{A}) Solar spectral irradiance (AM 1.5, global tilt), radiative heat flux of blackbody thermal radiation at various temperature, and reflectivity spectra of ideal SSAs and black surfaces. (\textbf{B}) and (\textbf{C}) The 3-D plot of solar to heat energy conversion efficiency for SSAs' contour plotted against concentration factor and cut-off wavelength at different operational temperature of 500$^{\circ}$C and 800$^{\circ}$C, respectively. (\textbf{D}) 3D schematic of proposed multilayer stack consisting of W, Al$_2$O$_3$, and SiO$_2$, the incidence angle, $\theta$, is defined as the angle between solar incident radiation and the surface normal. (\textbf{E}) A photo of sample fabrication on a 2 inch silicon wafer compared with a quarter coin and black painted paper. (\textbf{F}) A cross section SEM micrograph of the fabricated sample, the 2D schematic shows the thickness of each layer for the multilayer stack. (\textbf{F}) Normalized spectral distribution for radiative heat flux of solar (ASTM G173 AM 1.5) and blackbody thermal radiation (400$^{\circ}$C), as well as simulated and measured reflectivity spectrum of multilayer solar absorber. (\textbf{H}) The SSAs' high absorptance $\epsilon_{solar}$ $(\theta)$ and (\textbf{I}) $R_{IR}$ ($\theta$) across various angle of incident (AOI) result in excellent hemispherical $\epsilon_{solar}$ and $R_{IR}$. (\textbf{J}) The SSAs' high absorptance $\epsilon_{solar}$ $(\theta)$ and (\textbf{K}) $R_{IR}$ ($\theta$) across different angle of polarizations.
} 
\end{figure*}

According to the Planck's formula of blackbody radiation intensity distribution, we know the solar intensity and the blackbody intensity for mid temperatures (400$^{\circ}$C\textasciitilde700$^{\circ}$C) are distributed over different wavelength regimes (Fig. \ref{fig:sem_variangle}(\textbf{A})). The spectral irradiance of a 400$^{\circ}$C blackbody is at the same order of solar heat flux, while it will be twice as the solar intensity when the temperature of blackbody increases to 500$^{\circ}$C. Meanwhile, the expanded regime of blackbody radiation moves to a shorter wavelength as its temperature increases, as illustrated by peaks of the blackbody irradiance curves from 100$^{\circ}$C to 500$^{\circ}$C (Fig. \ref{fig:sem_variangle}(\textbf{A})). An ideal spectral selective absorber has zero reflectance at solar radiation regime and unity at the blackbody thermal radiation region, with a sharp transition between these two regions. The cut-off wavelength, $\lambda_{cut-off}$, at which the transition happens, is where the blackbody radiation begins to exceed the solar intensity. The ideal transition wavelength will shift to shorter wavelengths, called blue-shift, for higher-temperature SSAs. Consequently, it is a trade-off to design perfect SSAs at different engineering applications. The efficiency of an SSAs, $\eta_{abs}$, can be simplified as: $ \eta_{\mathrm{abs}}=\alpha_{\mathrm{abs}}-\epsilon_{\mathrm{abs}} (\sigma\left(T_{\mathrm{abs}}^{4}-T_{\mathrm{amb}}^{4}\right))/(C F \cdot Q_{\mathrm{abs}})$. Figure \ref{fig:sem_variangle}(\textbf{B}) and (\textbf{C}) are the 3D contour plots which illustrate the energy conversion efficiency, $\eta_{abs}$, as a function of the cut-off wavelength, $\lambda_{abs,cut}$ and concentration factors, CF, at different operational temperature, $T_{abs}$, of 500$^{\circ}$C, and 800$^{\circ}$C, in which the red color represents higher efficiency, and the blue means lower efficiency (plots by analyzing formulae section 1 in supplementary materials). The efficiency curves share the same trend of increasing to maximum from 0 and then decrease as $\lambda_{cut-off}$ sweeps from 0.25 $\mu$m to 4.0 $\mu$m (Fig. \ref{fig:sem_variangle}(\textbf{B}) and (\textbf{C})). The maximum efficiency of the 500$^{\circ}$C SSAs (Fig. \ref{fig:sem_variangle}(\textbf{B})), increases with the concentration factors varying from 1 to 100 at a fixed cut-off wavelength, which can be seen that the red color is getting deeper. The efficiency, $\eta_{abs}$, will increase when the concentration factor, CF, increases since the CF is in the denominator. Additionally, the corresponding cut-off wavelength will shift to a longer wavelength with the increasing of concentration factors, since the red-shift ensures that the SSAs absorb more solar energy. By comparing Fig. \ref{fig:sem_variangle}(\textbf{B}) and \ref{fig:sem_variangle}(\textbf{C}), it shows that the maximum energy efficiency of the same corresponding concentration factor drops with the increasing of operational temperature (black points in Fig. \ref{fig:sem_variangle}(\textbf{B}) and \ref{fig:sem_variangle}(\textbf{C})). It is reasonable that the energy efficiency, $\eta_{abs}$, decreases with the increasing of absorber operational temperature, $T_{abs}$, because $T_{abs}$ is in the numerator. 


\subsection{Characterizations of sample radiative properties}

\begin{figure*}[ht]
\centering
\includegraphics[width=1.0\textwidth]{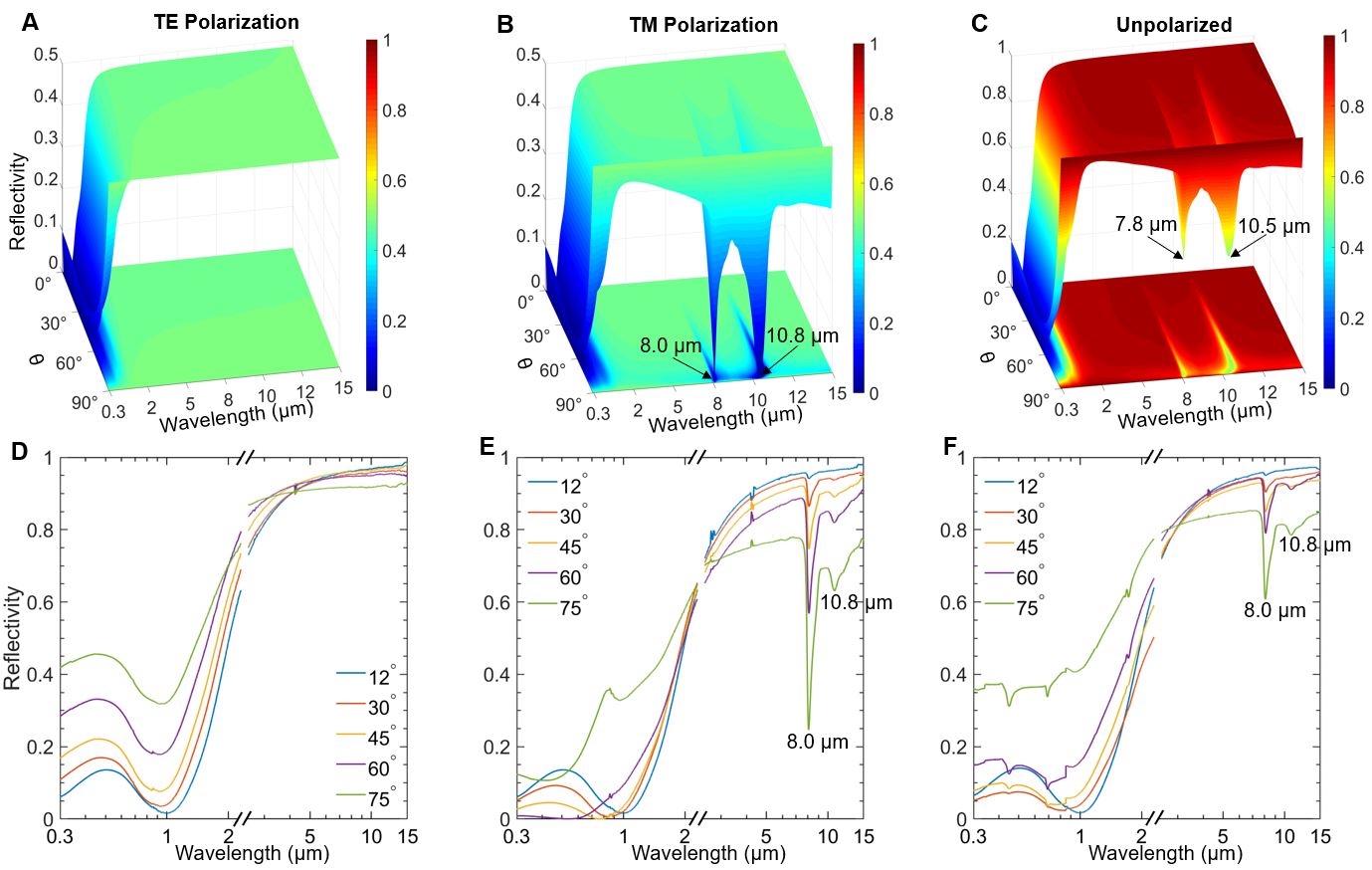}
\caption{ \label{fig:TE_TM} Simulated angle dependent TE polarization (\textbf{A}), TM polarization (\textbf{B}), and unpolarized (\textbf{C}) specular reflectivity spectra of SSAs' 3-D plot and projected 2-D contour plot as functions of wavelength and angle of incidence, $\theta$. Measured specular reflectivity spectra of SSAs across angles at TE polarization (\textbf{D}), TM polarization (\textbf{E}), and unpolarized light (\textbf{F}).
}
\end{figure*}

Figure \ref{fig:sem_variangle}(\textbf{D}) illustrates the 3D schematic of a proposed multilayered stack consisting of five layers. Figure \ref{fig:sem_variangle}(\textbf{E}) manifests that the fabricated SSA is black at various angles with visual observation, which elucidates its high absorptance in the visible wavelength. The hemispherical reflectivity which exhibits spectral selectivity with solar reflectivity $R_{abs}$ $<$ 0.06 in the solar spectrum and thermal reflectively $R_{abs}$ $>$ 0.94 in the mid-infrared region. 
The thickness of different layers for the fabricated sample is confirmed by the cross-section view of FE-SEM shown in Fig. \ref{fig:sem_variangle}(\textbf{F}). The bottom W layer is thick enough to block all the incident light and the solar absorptance of the absorber, $\alpha$ = 1 $-$ $R_{abs}$. Obviously, the solar irradiance randomly distributed in a one-day cycle, it is consequential to make the SSA facing to the sun all the time with the help of a dual-axis solar tracker system. The fabricated SSA has angular-independent hemispherical $\epsilon_{solar} (\theta)$ $>$ 0.90 (from 0.3$\mu$m to 2.5$\mu$m) (Fig. \ref{fig:sem_variangle}(\textbf{H})) and $R_{IR} (\theta)$ $>$ 0.93 (from 2.5$\mu$m to 20$\mu$m) (Fig. \ref{fig:sem_variangle}(\textbf{I}) (AOI, from 6$^{\circ}$ to 75$^{\circ}$), and it enables the SSAs latitude-insensitive and to work all day without solar-track systems.
In contrast, other SSAs report in literature often have directionally orientated structure, which enhances $\epsilon_{solar}, (0^{\circ})$, but reduces $\epsilon_{solar} (> 60^{\circ})$\cite{zhu2009optical,pettit1978solar}. As shown in Fig.\ref{fig:sem_variangle}({\textbf{H}}), for $\theta$ > 60$^{\circ}$, the SSAs have a higher $\epsilon_{solar}(\theta)$ than other SSAs. Besides, the hemispherical $\epsilon_{solar} (\theta)$ $>$ 0.89 (Fig.\ref{fig:sem_variangle}(\textbf{J}); AOI, 6$^{\circ}$) and $R_{IR} (\theta)$ $>$ 0.95 (Fig.\ref{fig:sem_variangle}(\textbf{K}); AOI, 12$^{\circ}$) for various polarization angles (from 6$^{\circ}$ to 75$^{\circ}$). Here, Figure \ref{fig:TE_TM} shows both the 3D contour plot of reflectivity spectra for the designed SSAs as functions of incident angle $\theta$, and wavelength $\lambda$, transverse electric (TE), transverse magnetic (TM), and unpolarized waves. The simulated reflectivity remains low within visible and near-infrared and high reflectivity ($>$ 0.95) in the mid-infrared region (from 3 $\mu$m to 15 $\mu$m) for unpolarized waves (Fig. \ref{fig:TE_TM}(\textbf{C})). Table S1 in supplementary materials lists the reflectivity at 0.55 $\mu$m, which the irradiance peak of solar lies, incident light at various angles. For both TE and TM waves, the reflectivity of the designed absorber is low from 0$^{\circ}$ to 75$^{\circ}$. However, the measured reflectivity of SSAs at 60$^{\circ}$ and 75$^{\circ}$ is higher than the simulated results, this is due to the fabrication process: the chemical bond (Si-O and Al-O) of SiO$_{2}$ and Al$_{2}$O$_{3}$ can be broken when collided with the charged energetic ions beams, resulting in the change of atomic ratio of Si/O and Al/O. Note that, the reflectivity spectra show two dips around $\lambda$ = 8.0 (7.8) $\mu$m and $\lambda$ = 10.8 (10.5) $\mu$m (Fig. \ref{fig:TE_TM}(\textbf{B}) and \ref{fig:TE_TM}(\textbf{C})), which are due to the surface phonon resonance of SiO$_2$ and Al$_2$O$_3$, respectively, this is also confirmed by the specular reflectivity measurements. However, the irradiance intensity peak of 500$^{\circ}$C blackbody thermal radiation lies at 3.8 $\mu$m (Fig. \ref{fig:sem_variangle}(\textbf{A})), while the blackbody emission power of 500$^{\circ}$C blackbody radiation at $\lambda$ = 8 $\mu$m and $\lambda$ = 11 $\mu$m only equals to 30\% and 11\% of the peak value. For higher temperature applications ($> 500^{\circ}$C), the blue-shifts of blackbody radiation will make the proportion of thermal radiation at around $\lambda$ = 8 $\mu$m and $\lambda$ = 11 $\mu$m even smaller and these dips are prominent only for higher incident angles, so it can be concluded that these two dips in reflectivity do not affect the selectivity performance of the SSAs. The overall absorptance of the fabricated samples at solar irradiance wavelength regime is 92\%, while it shows a low emittance of 6.6\% at a thermal wavelength with the measured spectral reflectivity under unconcentrated sunlight by solving Eq. S2 and S3 (the integral interval of solar radiation is from 0.25 $\mu$m to 2.5 $\mu$m, and the integral interval of blackbody thermal radiation is in the range of 2.5 $\mu$m - 15 $\mu$m).
\subsection{Ambient thermal resistance test and thermal degradation mechanism from SEM topography characterizations}

\begin{figure*}[!t]
\centering
\includegraphics[width=1.0\textwidth]{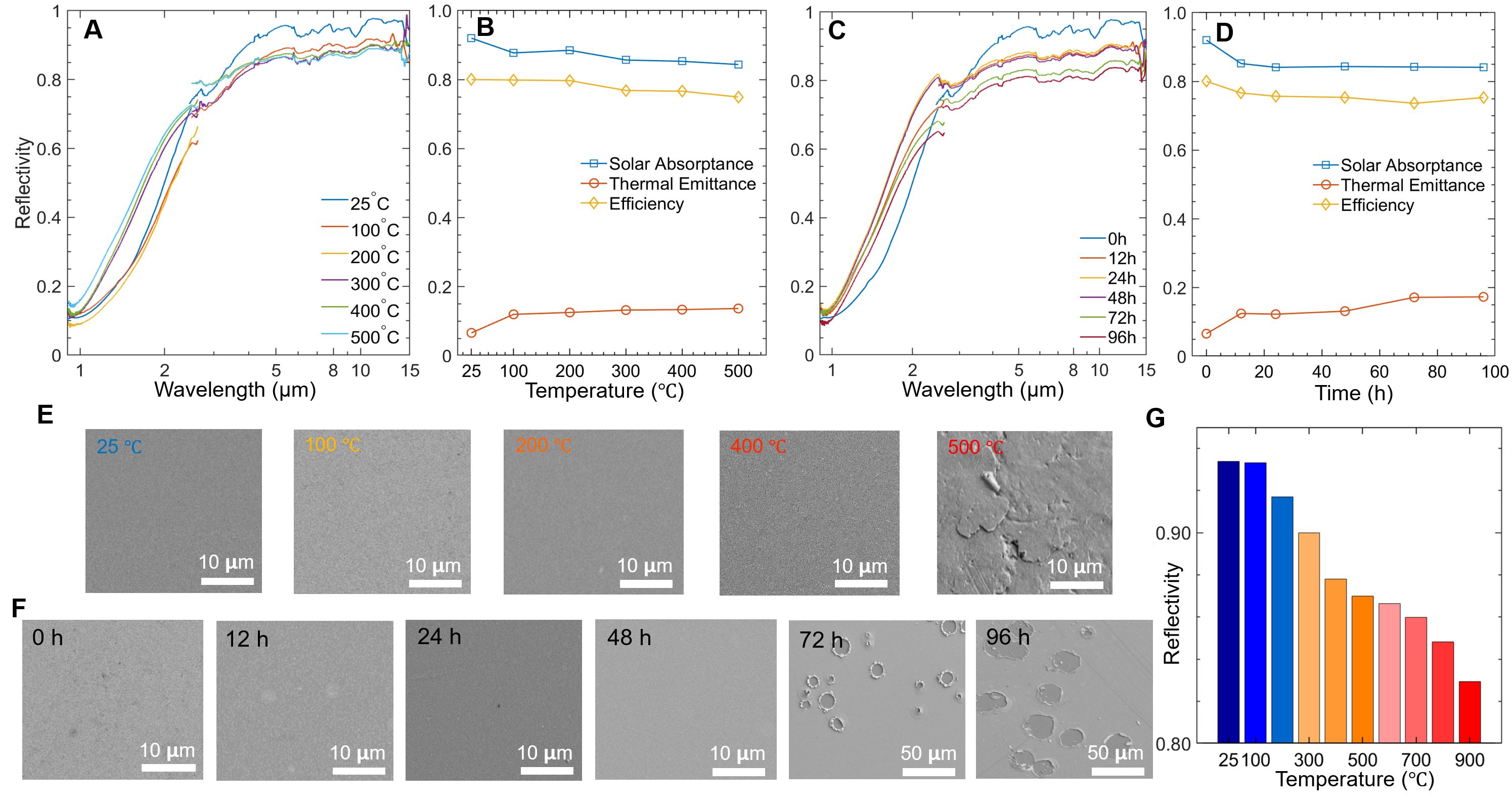}
\caption{\label{fig:thermaltest} Reflectivity spectra (\textbf{A}) and overall $\epsilon_{solar}$, $\epsilon_{IR}$, and $\eta_{abs}$ (\textbf{B}) of the SSAs measured by FTIR spectrometer after one-hour thermal treatment at different temperatures under ambient environment. Reflectivity spectra (\textbf{C}) and overall $\epsilon_{solar}$, $\epsilon_{IR}$, and $\eta_{abs}$ (\textbf{D}) of the SSAs at 400$^{\circ}$C after long-time thermal treatment. SEM topographic image of the fabricated SSAs sample as fabricated and after one-hour thermal annealing at various temperatures (\textbf{E}) and long-time thermal treatment (\textbf{F}). (The blackbody temperature is fixed to be 100$^{\circ}$ when calculate the $\eta_{abs}$) (\textbf{G}) Real-time measurement of $R_{IR}$ for SSAs at elevated temperature under high vacuum ($<$ 10$^{-3}$Pa).
}
\end{figure*}

To maintain high energy efficiency, the CSP system always works under the concentrated sunlight to obtain high operational temperature, so it is consequential to have a constant spectral performance for SSAs at high temperatures. For the sake of evaluating the radiative properties of SSAs after thermal annealing at elevated temperatures and long-time thermal resistance tests, we measure the reflectivity spectrum of the fabricated SSAs after each thermal test for one-hour thermal annealing (Fig. \ref{fig:thermaltest}(\textbf{A})) and long-time thermal cycle (Fig. \ref{fig:thermaltest}(\textbf{C})). Figure \ref{fig:thermaltest} (\textbf{B}) and (\textbf{D}) show the overall of $\epsilon_{solar}$, $R_{IR}$, and $\eta_{abs}$ of SSAs after elevated temperature and long-time thermal tests. The spectral reflectivity of the tested sample reduces from 25$^{\circ}$C to 200$^{\circ}$C (Fig. \ref{fig:thermaltest}(\textbf{A}) and (\textbf{B})), since the $\epsilon_{solar}$ decreases from 92.0\% to 88.5\% (95.3\% remain) and thermal emittance, $\epsilon_{IR}$ increase from 6.6\% to 12.53\% (89.8\% increase) and the $\eta_{abs}$ reduce by 3.31\%. Though the SEM topography of SSAs shows little change, the dielectric functions of SiO$_{2}$ and Al$_{2}$O$_{3}$ might change at elevated temperatures. The $\epsilon_{solar}$ and $\epsilon_{IR}$ keep changing from 200$^{\circ}$C to 500$^{\circ}$C, which results in a decrease in $\eta_{abs}$ from 75.7\% to 74.5\%. This is possibly caused by the SSAs' topography damage, as shown in Fig. \ref{fig:thermaltest}(\textbf{E}) for 400$^{\circ}$C and 500$^{\circ}$C. Additionally, the cut-off wavelength of SSAs shifts to shorter wavelengths to maintain a high efficiency at high temperatures, as discussed in Sec \ref{sec2}.1.

Figure \ref{fig:thermaltest}(\textbf{C}) elucidates the reflectivity spectra of the fabricated samples after long-time thermal resistance tests. The reflectivity spectrum of SSAs changes severely after 400$^{\circ}$C thermal resistance test over 12h and 
$\epsilon_{solar}$ decreases from 92.0\% to 85.2\% (92.6\% remain) and thermal emittance $\epsilon_{IR}$ increases from 6.6\% to 12.5\% (89.8\% increase) and $\eta_{abs}$ only reduces by 3.39\%. After that, from 12h to 48h,  $\epsilon_{solar}$ and $\epsilon_{IR}$ barely change (Fig.\ref{fig:thermaltest}(\textbf{D})). However,  $\epsilon_{IR}$ increases from 48h to 72h and causes a drop of $\eta_{abs}$.

To understand the mechanism that causes the degradation above 400$^{\circ}$C and after 72h thermal treatment, the samples are characterized using FE-SEM (SIGMA VP). Figure \ref{fig:thermaltest}(\textbf{E}) shows the SEM topographic images of the SSAs surface before and after thermal annealing for one hour. From 25$^{\circ}$C to 400$^{\circ}$C, it is difficult to see any apparent changes. When the temperature keeps going up from 400$^{\circ}$C to 500$^{\circ}$C, small show up, which is also demonstrated in the reflectivity spectra in Fig. \ref{fig:thermaltest}(\textbf{A}), due to the thermal stress coming from the difference in thermal expansion coefficient between SiO$_2$ (0.55 $\times$ 10$^{-6}$ m/m$\cdot$K) and W (4.2 $\times$ 10$^{-6}$ m/m$\cdot$K) \cite{chirumamilla2016multilayer}. Figure \ref{fig:thermaltest}(\textbf{F}) shows the SEM topographic images of the SSAs surface before and after long-time thermal treatment. Little topographic change of SSAs can be seen from 12h to 48h, while small round cracks like burst bubbles appear after 72h and become bigger after 96h. It agrees with the reflectivity spectra change of Fig. \ref{fig:thermaltest}(\textbf{C}). It provides a guideline for improvement approaching a perfect SSAs by the utilization of other materials with a similar thermal expansion coefficient to avoid the thermal stress at high temperature and long-time heating. 

Figure \ref{fig:thermaltest}(\textbf{G}) shows that the $R_{IR}$ of SSAs gradually drops heating under elevated temperatures. Each stepped drops from 100$^{\circ}$C to 400$^{\circ}$C is larger than that from 400$^{\circ}$C to 800$^{\circ}$C. It proves that most of the thermal stress between different layers of SSAs is released when heated from 100$^{\circ}$C to 400 $^{\circ}$C.
Combined with the SEM topographic image of Fig. \ref{fig:thermaltest}(\textbf{E}), we conclude that the formation of cracks contributes more than the development of cracks to the decrease of $R_{IR}$. Even heated up to 900$^{\circ}$C, the $R_{IR}$ of SSAs still remain 90.15\% compared with the value at ambient temperature, this indicates that the combinations of SiO$_{2}$ and Al$_{2}$O$_{3}$ has the potential applications under high-temperature solar thermal systems.

\begin{figure}[!ht]
\centering
\includegraphics[width=1.0\textwidth]{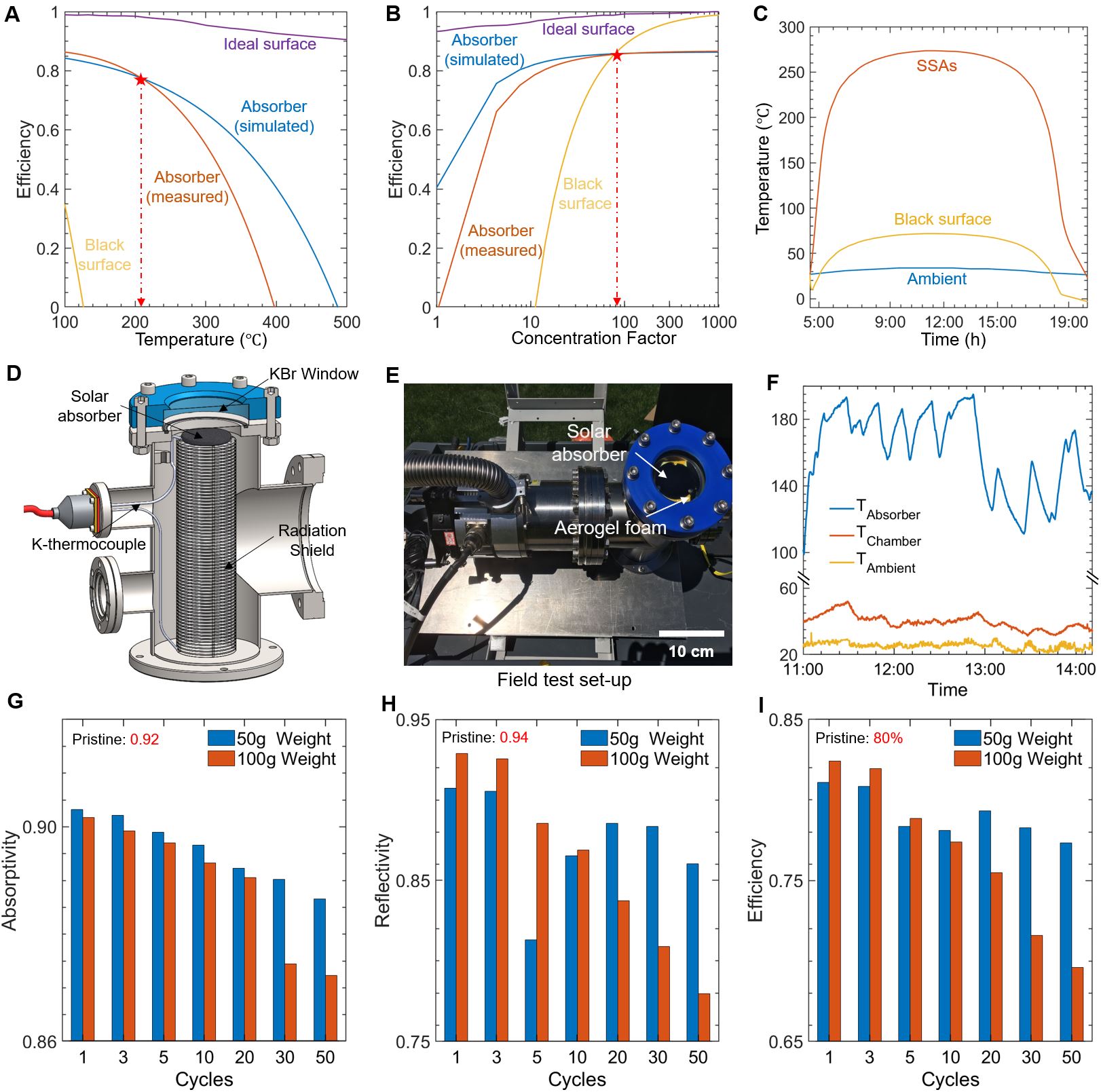}
\caption{ \label{fig:eff_mechanical_field} (\textbf{A}) Solar to heat conversion efficiency as a function of absorber operational temperature, $T_{abs}$, for an ideal selective absorber, SSAs with reflectivity spectrum of measured or simulated, and a black surface, under the unconcentrated solar light (1 sun); (\textbf{B}) Solar to heat conversion efficiency for abovementioned four surfaces as a function of concentration factor, CF, at the operational temperature of $T_{abs}$ = 400$^{\circ}$C. (\textbf{C}) Thermal performance of the selective absorber (orange curve) and the black surface (yellow curve) over a one-day solar cycle from sunrise (5:00 a.m.) to one hour after sunset (8:00 p.m.) at varying ambient temperature (blue curve) under 1 sun. 3D schematic (\textbf{D}) of the real (\textbf{E}) outdoor experimental set-up. (\textbf{F}) The temperature variations of the sample's surface, vacuum chamber, and outdoor ambient environment. $\epsilon_{solar}$ (\textbf{G}), $R_{IR}$ (\textbf{H}), and $\eta_{abs}$ (\textbf{I}) after the different abrasion tests under different weight (50g/100g) as a function of cycles. 
}
\end{figure}

\subsection{Solar conversion efficiency investigations of fabricated multilayer SSAs}

As discussed above, the designed structure of SSAs is demonstrated to be angular-insensitive as well as thermal stable up to 400$^{\circ}$C. We employ the reflectivity measured by the UV/VIS/NIR spectrophotometer and FTIR spectrometer at room temperature (25$^{\circ}$C), as shown in Fig. \ref{fig:sem_variangle}(\textbf{G}). The $\alpha_{\lambda,abs}^{\prime}$ and $\epsilon_{\lambda,abs}^{\prime}$ are assumed to be independent of temperature as observed in Fig. \ref{fig:thermaltest}(\textbf{A}). The spectral integration for $\alpha_{\lambda,abs}^{\prime}$ and $\epsilon_{\lambda,abs}^{\prime}$ is performed over wavelengths from 0.3 $\mu$m to 16 $\mu$m, which covers 99.9\% of the solar radiation and only 4.7\% of thermal radiation energy outside this spectral region for a 400$^{\circ}$C blackbody. The reflectivity of the ideal absorber is zero below the cut-off wavelength, while its reflectivity is unity beyond the cut-off wavelength to minimize thermal leakage from blackbody radiation (Fig. \ref{fig:sem_variangle} (\textbf{A}), the dark-orange dot-dash line). The cut-off wavelength is optimized according to the shifts of blackbody thermal radiation at different operational temperatures to maximize the energy conversion efficiency, which defines an upper limit of system efficiency. The reflectivity of black surface, $R_{abs,black}$, is zero over the entire wavelength region ($\alpha$ = 1 -- $R_{abs,black}$ = 1) and shows no spectral selectivity (Fig. \ref{fig:sem_variangle}(\textbf{A}), the black long dash line).

The conversion efficiencies (by analyzing formulae section 1 in supplementary materials) of SSAs are 84.28\% and 86.32\% for the simulated and measured spectral reflectivity at $T_{abs}$ = 100$^{\circ}$C, respectively (Fig. \ref{fig:eff_mechanical_field}(\textbf{A})). The efficiency drops gradually to zero at the stagnation temperature of 396$^{\circ}$C and 486$^{\circ}$C using simulated and measured optical properties, respectively. The absorbed solar energy equals the blackbody re-emission energy (i.e., no solar thermal energy is actually harvested) at this point. The stagnation temperature of SSAs using the simulated reflectivity spectrum is higher than the one using the measured reflectivity spectrum since the reflectivity curve of the simulated one is steeper than the measured one over the transition wavelength region. The efficiency curves of SSAs with measured and simulated radiative properties intersect at 210$^{\circ}$C. The fabricated SSA has a higher efficiency than the designed SSA below 210$^{\circ}$C, because the measured spectrum has a higher reflectivity than the simulated one from 7.7 $\mu$m to 20 $\mu$m, (as marked by red solid pentagram in Fig. \ref{fig:sem_variangle}(\textbf{G})). Half of the blackbody thermal radiation spreads in this wavelength region (the thermal radiation of a 210$^{\circ}$C blackbody is within 3.0 $\mu$m to 15 $\mu$m). When the temperature goes above 210$^{\circ}$C, the efficiency of SSAs using simulated reflectivity spectrum exhibits an advantage than the measured one since the blackbody radiation blue-shifts to the shorter wavelength, where the simulated spectrum has a higher reflectivity than the measured. 
As a reference, the black surface can only convert 34.8\% solar energy to heat at $T_{abs}$ = 100$^{\circ}$C, and its efficiency goes down to zero at 126$^{\circ}$C, which further demonstrates the significance of spectral selectivity in enhancing the solar to heat conversion efficiency. On the other hand, the efficiency of the multilayered absorber is 17\% and 20\% lower than the ideal surface at $T_{abs}$ = 100$^{\circ}$C with measured optical properties and simulated radiative spectrum data, respectively. This gap becomes large when the stagnation temperature increases. It mainly results from the larger thermal emittance (around 0.09) in the mid-infrared regime and the cut-off wavelength of the ideal surface is optimized at each temperature according to Wien's displacement law, while the cut-off wavelength of selective absorber keeps unchanged at around 1.2 $\mu$m. Simultaneously, the reflectivity spectrum of the ideal surface changes much sharply at the cut-off point than the multilayered SSAs, comparing Fig. \ref{fig:sem_variangle}(\textbf{A}) and Fig. \ref{fig:sem_variangle}(\textbf{G}). Therefore, the geometry parameters of the multilayered SSAs need to be optimized to make the cut-off wavelength perfectly matched the operational temperature. 

Figure \ref{fig:eff_mechanical_field}(\textbf{B}) shows the efficiency as a function of concentration factors, CF, from 1 to 1,000, when $T_{abs}$ = 400$^{\circ}$C, aiming at a mid-temperature application. The cut-off wavelength of the ideal surface is optimized according to different concentration factors. This indicates an upper limit for the SSAs' performance under different concentration factors. The energy efficiency of SSAs with measured or simulated optical properties and the black surface keeps going up with the increasing of CF. The efficiency of SSAs with the measured reflectivity spectrum is lower than the simulated one because the simulated spectrum has a higher solar absorptance than the measured one. For the black surface, its energy conversion efficiency becomes greater than zero at around 12 suns and climbs close to the ideal surface when the CF = 1,000 since the solar radiation heat flux is much larger than the 400$^{\circ}$C blackbody thermal radiation under 1000 suns.

\subsection{Thermal performance investigations under unconcentrated solar radiation}

Figure \ref{fig:eff_mechanical_field}(\textbf{C}) shows the calculated transient temperature variations of SSAs and the black surface under one sun over a solar cycle in Boston, Massachusetts \cite{Weatherboston} using the ambient temperature \cite{Weatherboston} and the solar illumination data \cite{solarangleboston} of July 10, 2018 (see formulae section 2). The highest temperature of the SSAs is 273$^{\circ}$C, while the highest temperature of the black surface is 72$^{\circ}$C. The thermal performance of the selective solar absorber is overwhelmingly better than the black surface at any time under sunshine. The highest temperature difference between SSAs and the black surface is 311$^{\circ}$C under 20 suns (see supplementary Fig. S1), which is higher than the temperature difference under one sun (201$^{\circ}$C). We evaluate the outdoor test performance by measuring the stagnation temperature under unconcentrated solar radiation in Kingston, Rhode Island on May 27, 2019. A SSAs sample was placed in a 100 cm diameter vacuum chamber which was equipped with a Potassium bromide (KBr) window (70 cm in diameter) (Fig. \ref{fig:eff_mechanical_field}(\textbf{D}) and (\textbf{E})). The KBr window is highly transparent (\textgreater 93\%) from UV to the mid-infrared range, which allows sunlight to come in and the thermal radiation of the heated SSAs re-emit to the sky. The vacuum chamber was set on a mounting frame with an angle of $\sim$ 32$^{\circ}$ to ensure the absorber was normal to the incoming sunlight. A K-type thermocouple was secured to the back of the SSAs sample with Kapton tape. Another K-type couple was taped to the bottom of the radiation shield to measure the local temperature of the vacuum chamber. In order to eliminate the conductive and radiative heat losses, we supported the absorber sample with low thermal conductivity (29 mW/m$\cdot$K) aerogel tiles that were on top of the radiation shield made of several double-side polished aluminum sheets. The vacuum chamber was pumped continuously to be $\sim$ 5 $\times$ 10$^{-3}$ Pa with the turbomolecular pump to eliminate the convective and conductive thermal losses from the air. The warmed up to the maximum stagnation temperature of 193.5$^{\circ}$C at around 11:30 since the weather was partly cloudy, the temperature of the vacuum chamber and the absorber sample showed strong fluctuations during the test and was lower than the calculated maximum temperature. The stagnation temperature is enough for the rooftop solar thermal systems even under partly cloudy weather. 

\subsection{SSAs’ surface abrasion robustness tests}
Figure \ref{fig:eff_mechanical_field}(\textbf{G}, \textbf{H}, and \textbf{I}) show $\epsilon_{solar}$ (\textbf{G}), $R_{IR}$ (\textbf{H}), and $\eta_{abs}$ (\textbf{I}) of SSAs after each abrasion cycle (the definition of one abrasion cycle is in supplementary Fig. S2 ). It can be found that the $\epsilon_{solar}$, $R_{IR}$, and $\eta_{abs}$ remain 94.6\%, 83.5\%, and 86.9\% after 50 abrasion cycles under 100g weight (2040 Pa). The $R_{IR}$ of SSAs under 50g weight (1020 Pa) first drops and then increases, since the pressure of 50g weight can remove part of the top deposited SiO$_2$ and Al$_2$O$_3$ layer and make the bottom W layer exposed out in 5 cycles. While the 100g weight can make part of the bottom W layer exposed out only in 1 cycle. Since the abrasion resistance tests will break the fabricated SSAs layer, the $\epsilon_{solar}$ keeps decreasing whichever 50g or 100g weight is placed above. The $\eta_{abs}$ of 50g weight tests first decreases and goes up after 10 cycles, due to the $R_{IR}$ fluctuations of 50g weight tests.

\section{Discussion}
\label{Sec3}
The spectral selectivity and thermal performance of multilayered SSAs consisting of Al$_2$O$_3$-W-SiO$_2$-W stacks are investigated analytically and experimentally. It demonstrates 92\% solar absorptance and 6\% reflectivity in the blackbody thermal radiation region. Oblique reflectivity is also experimentally characterized to show high spectral selectivity for both TE and TM polarizations, which demonstrates its angular and polarization insensitivity. High-temperature thermal annealing performed at elevated temperatures for one hour shows the fabricated SSAs maintain its wavelength selectivity at even 400$^{\circ}$C and 500$^{\circ}$C. The high-temperature thermal resistance test is investigated under 400$^{\circ}$C for 12h, 24h, 48h, 72h, and 96h cycles and it demonstrates SSAs keep good selectivity within 48h, while slight degradation of spectral selectivity happens after the 72h and 96h thermal treatment. Real-time reflectivity measurements show that the material displays only a minor change in optical properties above 800$^{\circ}$C.  Our study indicates that the proposed metamaterial exhibits thermal resistance and practically retains its optical proprieties even after 96h operation at 400$^{\circ}$C. Retention of optical properties after high temperature annealing can be attributed to pattern-free design of the material as cracks can be significantly more detrimental to optical properties of surface designs such as gratings. The outdoor tests show a peak stagnation temperature of 193.5$^{\circ}$C, and thus indicate its tremendous potential for low temperature solar thermal applications in the absence of solar concentrator. Additionally, the surface abrasion test yields that SSAs have a robust resistance to sandpaper abrasion attack for a long-duration practical application, Abrasion resistance can also be attributed to this pattern-free design.

\section{Materials and Methods}
\label{Sec4}
\subsection{Materials}
SiO$_2$ (99.995\%, 4N5), Al$_2$O$_3$(99.995\%, 4N5), and W (99.995\%, 3N5) are purchased from Angstrom Sciences, Inc. 2 inches wafers (Type: P; dopant: B; orientation: <1-0-0>; resistivity: 10-20 ohm/cm; thickness: 279$\pm$25 $\mu$m) are purchased from WaferPro.
\subsection{Methods}
\subsubsection{Sample fabrication and SEM topography characterizations}
SSAs samples are deposited with a magnetron sputtering technique within a home-built high-vacuum sputtering system (see supplementary Fig. S3). The fabrication process has been described in detail in this publication \cite{tian2019near}. The fabrication parameters, such as deposition rate, base pressure, and sputtering power for each deposition procedure are specified in Table S2 in supplementary materials. The base pressure before sputtering is 3.4 $\times$ 10$^{-7}$ Torr. The Al$_2$O$_3$ and W layer are both deposited directly from Al$_2$O$_3$ and W target, respectively, while the SiO$_2$ layer is oxidized from the silicon (Si) with radio frequency (RF) sputtering. The cross-section and topography of absorber samples after various temperature thermal annealing and diverse long-time thermal resistance tests are characterized by SIGMA VP Field Emission-Scanning Electron Microscope (FE-SEM).

\subsubsection{Optical properties measurements}
Hemispherical reflectance measurements in the UV, visible, and near-infrared regions (from 0.3 $\mu$m to 2.5 $\mu$m) are performed on a Jasco V770 spectrophotometer with a 60 mm PTFE based integrating sphere. A wavelength scan step is fixed as 1 nm with AOI of 6$^{\circ}$ and normalized to a labsphere spectralon reflectance standard. The mid-infrared (2.5 $\mu$m to 15 $\mu$m) reflectance measurements are executed on Jasco 6600 FTIR spectrometer together with a Pike 3-inches golden integrating sphere. The AOI of the incident beam from the FTIR spectrometer is fixed at 6$^{\circ}$ and the diffused golden reference is selected as a normalized standard. These spectra are taken at a scan rate of 64 with a wavelength resolution of 0.4 cm$^{-1}$. Details about FTIR spectrum measurements can be referred to the recent article \cite{tian2019near}. 

The hemispherical reflectance measurement for different angles of incident (AOI) is conducted by using home-built wedge blocks (see supplementary Fig. S4) with (18$^{\circ}$, 30$^{\circ}$, 33$^{\circ}$, 33$^{\circ}$, 51$^{\circ}$, 66$^{\circ}$, and 76$^{\circ}$)  combined with Jasco 60mm PTFE based integrating sphere and Pike 3-inches golden integrating sphere. The wedge blocks for PTFE integrating sphere are put faced to the sample ports of the Jasco and Pike integrating sphere. SSAs samples and the reference standard are put faced to the sample ports on designed wedge blocks. Background spectrum and the reflectance spectrum are taken as usual.

The specular reflectance measurements of variable angle in the UV, visible, and near-infrared regions (from 0.3 $\mu$m to 2.5 $\mu$m) are performed on Jasco V770 spectrophotometer with Harrick variable angle reflection accessory. The variable angle specular reflectance measurements (from 2.5 $\mu$m to 15 $\mu$m) are executed on Jasco 6600 FTIR spectrometer together with Harrick Seagull$^{TM}$ Variable Angle Reflection Accessory. Both measurements are taken for angles of  12$^{\circ}$, 30$^{\circ}$, 45$^{\circ}$, 60$^{\circ}$, and 75$^{\circ}$. 

The hemispherical reflectance measurements of variable angle polarization from 0.3 $\mu$m to 2.5 $\mu$m are done on Jasco V770 spectrophotometer with GPH-506 polarizer and 60 mm integrating sphere. The hemispherical reflectance measurements of variable angle polarization from 2.5 $\mu$m to 15 $\mu$m are executed on Jasco 6600 FTIR spectrometer together with PL-82 polarizer and Pike 3-inches golden integrating sphere.

\subsubsection{High-temperature thermal annealing and resistance tests}
High-temperature thermal annealing and resistance tests are both carried out in a tube oven at an ambient atmosphere with an alumina tube of 5 cm diameter and 80 cm length. Samples are put in an alumina crucible boat (100 mm $\times$ 30 mm $\times$ 20 mm) that is placed in the center of the tube. The increasing rate of temperature is set to be at 15$^{\circ}$C/min for 100$^{\circ}$C 200$^{\circ}$C 300$^{\circ}$C 400$^{\circ}$C, and 500$^{\circ}$C for one hour. The hemispherical reflectance measurements are carried out after each thermal treatment. For the high-temperature thermal resistance tests, the temperature controller of the tube oven is set to be 400$^{\circ}$C for 12h, 24h, 48h, 72h, and 96h thermal cycles. The samples are taken out after each thermal cycle for reflectance spectrum measurements and a small piece of the sample ($\sim$ 3mm $\times$ 3mm) is cut for SEM topography characterization.

\subsubsection{Real-time reflectivity measurement of SSAs under high vacuum}
The real-time measurement of reflectivity for SSAs is conducted on Jasco FTIR 6600 spectrometer within the Harrick high-temperature reaction chamber. The chamber can be heated by an electrical heater up to 910 
$^{\circ}$C and vacuumed down to < 10$^{-3}$ Pa connected to a Pfeiffer TMH 260 turbopump. The temperature of the SSAs is measured by a K-type thermocouple and controlled by the Waterloo PID temperature controller.

\subsubsection{SSAs’ surface abrasion robustness tests}
A sandpaper abrasion test was carried out using 400 grit SiC sandpaper as an abrasion surface, as shown in supplementary Fig. S2. The SSAs samples with a weight of 50/100 g above it are put face-down to sandpaper and move 10 cm along the ruler (see supplementary Fig. S2 B); then, the sample is then rotated counterclockwise by 90$^{\circ}$ (face to the sandpaper) and move 10 cm along the ruler. The three steps shown are in supplementary Fig. S2 B is defined as one abrasion cycle. This guarantees that the SSAs’ surface is abraded longitudinally and transversely in each cycle. The reflectivity spectra are measured after 1, 3, 5, 10, 20, 30, and 50 cycles.

\section{Acknowledgements} This project is supported in part by the National Science Foundation through grant numbers CBET-1941743, and National Aeronautics and Space Administration through grant number NNX15AK52A.

\section{Author Contributions}
Y.T. and X.L. develop the theoretical calculations and carry out spectrum measurements and field tests. L.Q. fabricates SSAs samples. A.G. and J.L. contribute to the interpretation of the results. All authors provide critical feedback and help revise the final version of the manuscript. T.T., G.X., and Y.Z. supervise the project. 

\section{Conflicts of interest}
The authors declare no conflict of interest.

\section{Bibliography}








\end{document}